\documentstyle[graphicx,prl,multicol,amsmath,aps,epsfig]{revtex}

\begin{document}

\draft
\sf

\title{Comparison of Josephson vortex flow transistors with different gate line configurations}

\author{\renewcommand{\thefootnote}{\alph{footnote})}
J.~Schuler,\footnote{\baselineskip4mm present address: Walther-Mei{\ss}ner-Institut,
Bayerische Akademie der Wissenschaften, Walther-Mei{\ss}ner
Str.~8, D-85748 \hspace*{5mm}Garching, Germany}
S.~Weiss, T.~Bauch, A.~Marx,$^{a)}$
D.~Koelle,\footnote{e-mail: koelle@ph2.uni-koeln.de}
and R.~Gross$^{a)}$}

\address{II.~Physikalisches Institut, Universit\"{a}t zu K\"{o}ln, Z\"{u}lpicher
Str.~77, D - 50937 K\"{o}ln, Germany}

\maketitle


\begin{abstract}
We performed numerical simulations and experiments on Josephson
vortex flow transistors based on parallel arrays of $\rm YBa_2Cu_3O_{7-\delta}$
grain boundary junctions with a cross gate-line allowing
to operate the same devices in two different modes named Josephson fluxon
transistor (JFT) and Josephson fluxon-antifluxon transistor (JFAT).
The simulations yield a general expression for the current gain
vs. number of junctions and normalized loop inductance
and predict higher current gain for the JFAT.
The experiments are in good agreement with simulations and show improved
coupling between gate line and junctions for the JFAT as compared to the JFT.
\end{abstract}






\begin{multicols}{2}


The availability of a superconducting transistor is expected to
significantly expand the application range of superconducting electronics.
This has stimulated considerable activities towards the development of
three-terminal devices based on Josephson junctions from
high-transition-temperature superconductors (HTS), which offer high
intrinsic speed due to large products of critical current times junction
resistance. Among the various types which have been investigated, the
Josephson vortex flow transistor (JVFT) shows  probably the most promising
performance \cite{gross95a}. In JVFTs the density of Josephson vortices in
either a long Josephson junction or a parallel array of $N$ small junctions
is controlled by applying a magnetic field via a control (gate) current
$I_g$ flowing through a gate line. Hence, the critical current $I_c$ or the
voltage $V$ across the junction(s), if biased at $I_b>I_c$, is controlled by
$I_g$. A figure of merit is the current gain $g = | dI_c / dI_g |_{max}$,
which is defined as the slope of the $I_c(I_g)$ characteristics in its
steepest point. To maximize the current gain various JVFT designs have been
developed which basically differ in (i) the {\it type of Josephson
junctions}, i.e. either a single long junction or an array of $N$ short
junctions, coupled via $N$-1 loops (ii) the {\it junction geometry}, e.g.
symmetric overlap or asymmetric in-line type and (iii) the {\it gate line
geometry}, in particular the relative orientation of the gate line with
respect to the junction(s).

\begin{figure}[b]                                               %
\center{\includegraphics [width=0.95\columnwidth] {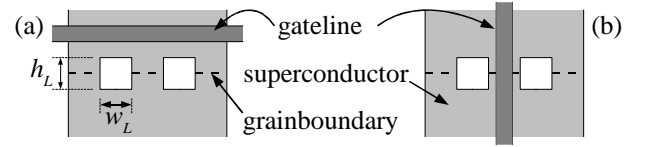}} %
\caption[]{\small Sketch of a JFT (a) and a JFAT (b) based on a %
discrete array of HTS grain boundary Josephson junctions.       %
\label{f-bauform}}                                              %
\end{figure}

We will focus here on parallel arrays of Josephson junctions which offer
more flexibility in design parameters as compared to long junction JVFTs
(e.g. variation of loop inductance $L$ via the loop dimensions). In a
standard geometry the gate line is positioned parallel to the array, as
shown in Fig.\ref{f-bauform}(a). We refer to this configuration as the {\it
Josephson fluxon transistor (JFT)}. In this design the current gain can be
drastically improved by introducing an asymmetric junction geometry
\cite{gross95a,gerdemann95}. Alternatively, instead of altering the bias
current distribution, the gate line can be oriented perpendicularly to the
junctions in the {\it Josephson fluxon-antifluxon transistor (JFAT)}, shown
in Fig.\ref{f-bauform}(b) \cite{berkowitz96,terzioglu96}. The JFAT seems to
show improved gain as compared to the symmetric JFT \cite{berkowitz96}. In
this letter we present a comparative study of both designs by numerical
simulation and by experiment, which is intended to clarify their different
behavior and which has not been understood so far.


To model both the JFT and JFAT we numerically
calculated the $I_c(I_g)$-characteristics for both gate
line geometries by using the image current model
\cite{terzioglu96} with an algorithm as described in
detail in Ref.~5\nocite{schuler98}.
In brief, the algorithm calculates the distribution of
junction currents and transverse currents in a network
of $N$ parallel junctions coupled by $N-1$ loops as a
function of applied magnetic flux and bias current
distribution in a self consistent way. We use the first
Josephson equation and fluxoid quantization for each
Josephson junction and each loop and current
conservation at each node of the array. Solutions for
the maximum total supercurrent through the network
$I_c$ are obtained by an iteration method which is
similar to the one developed for long junctions as
described in \cite{mayer95}.
The gate line is assumed to be separated from the bottom
superconducting film by a thin insulating layer, and a gate current $I_g$
induces an image current $I_i= - k_i I_g$ in the superconductor. Obviously,
the gain scales linearly with the coupling constant $0 \leq k_i \leq 1$, and
for the simulated $I_c(I_g)$-characteristics we assume $k_i = 1$. All
simulation results presented here were obtained with the assumption of
homogeneous distributions of the bias current $I_b$, the critical junction
currents $I_c^0$ and the inductance $L$ of the loops with a fixed ratio $w_L
/ h_L = 2$ of loop width $w_L$ and loop height $h_L$. To a first
approximation the value of $w_L / h_L$ does only affect the coupling
coefficient. Hence, all results regarding the scaling of $g(N,\beta_L)$ are
not affected by the choice of $w_L / h_L$.

\begin{figure}[tbh]                                                        %
\center{\includegraphics [width=0.8\columnwidth] {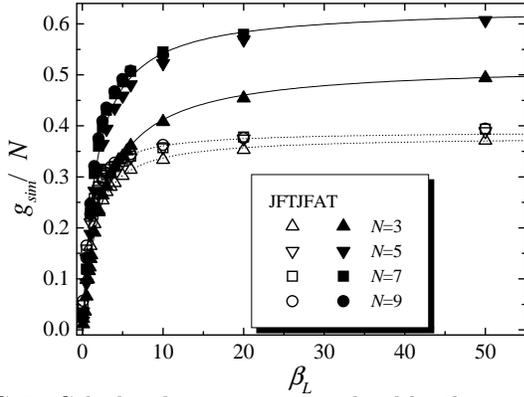}}               %
\caption[]{\small Calculated gain $g_{sim}$ normalized by the number of
junctions $N$ vs. normalized loop inductance $\beta_L$ for a JFT and a JFAT.
The lines are calculated from eq.(\ref{e-fit}).
\label{f-gsim}}                                                            %
\end{figure}                                                               %
From the numerically calculated $I_c(I_g)$-characteristics we extract the
current gain which depends on the type of gate configuration, number $N$ of
junctions in the array and on the normalized loop inductance $\beta_L = 2
\pi L I_c^0 / \Phi_0$, where $\Phi_0$ is the flux quantum. Figure
\ref{f-gsim} summarizes our simulation results in a plot of gain per
junction $g_{sim} / N$ vs. $\beta_L$ for the JFT (open symbols) and the JFAT
(solid symbols) for different values of $N$. For $N \geq 5$ the gain
$g_{sim}$ is linear in $N$ for both gate line configurations. Only in the
case of $N=3$ the gains per junction are noticeably lower. Furthermore,
$g_{sim} / N$ increases monotonically with $\beta_L$ and saturates at
$g_{sim} / N = g_\infty$ if $\beta_L\gg\beta_0$. Finally, $g_{sim} / N$ is
higher for the JFAT than for the corresponding JFT, except for small values
$\beta_L \leq 2.4$ (for $N=3$) and $\beta_L \leq 0.6$ (for $N\geq 5$). In
any case, $g_{sim}(N,\beta_L)$ can be very well fitted by

\begin{equation}
g_{sim} = N g_\infty \frac{\beta_L}{\beta_L + \beta_0} \quad ,
 \label{e-fit}
\end{equation}

 \noindent
with the fit parameters $g_\infty$ and $\beta_0$ given in table I.

\begin{figure}[t]                                                     %
\center{\includegraphics [width=0.8\columnwidth] {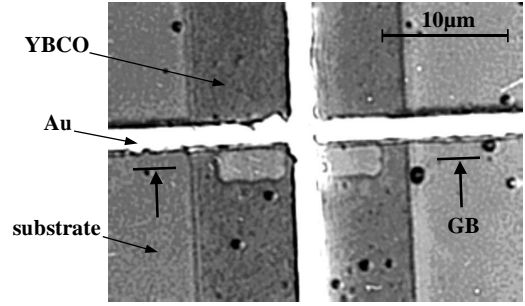}}         %
\caption[]{\small Optical micrograph of a JVFT with $N = 3$ YBCO grain boundary junctions, and
an Au cross gate line, which allows operation of the device either as JFT or as JFAT. The position of
the grain boundary is marked as GB.                             %
\label{f-photo}}                                                      %
\end{figure}

To compare our numerical simulation results with the experiment we
fabricated JVFTs based on $\rm YBa_2Cu_3O_{7-\delta}$ (YBCO) grain boundary
Josephson junctions. To place the gate line directly over the junction
arrays we used a three layer process, with the YBCO film as the bottom layer
separated by a polyimide insulating layer from the Au film forming the gate
lines. The 100\,nm thick YBCO was epitaxially grown by pulsed laser
deposition on a symmetric 24$^\circ$ SrTiO$_3$ bicrystal and subsequently
patterned using standard photolithography and Ar ion beam milling. Then, the
sample was spin coated with polyimide and after a soft-bake the polyimide
was photolithographically patterned and post-baked. Finally, the top 50\,nm
Au layer was deposited by e-beam evaporation using a lift off process to
form the gate lines. All fabricated devices discussed in this paper have
loop dimensions $w_L = 6 \mu$m and $h_L = 3 \mu$m, and 1-2$\mu$m wide
junctions. Figure \ref{f-photo} shows a device with $N=3$ junctions. Our
design with a cross gate configuration permits two modes of operation either
as a JFT or as a JFAT. This allows direct comparison of both transistor
geometries within a {\it single device}.

\begin{figure}[bth]                                                   %
\center{\includegraphics [width=0.8\columnwidth] {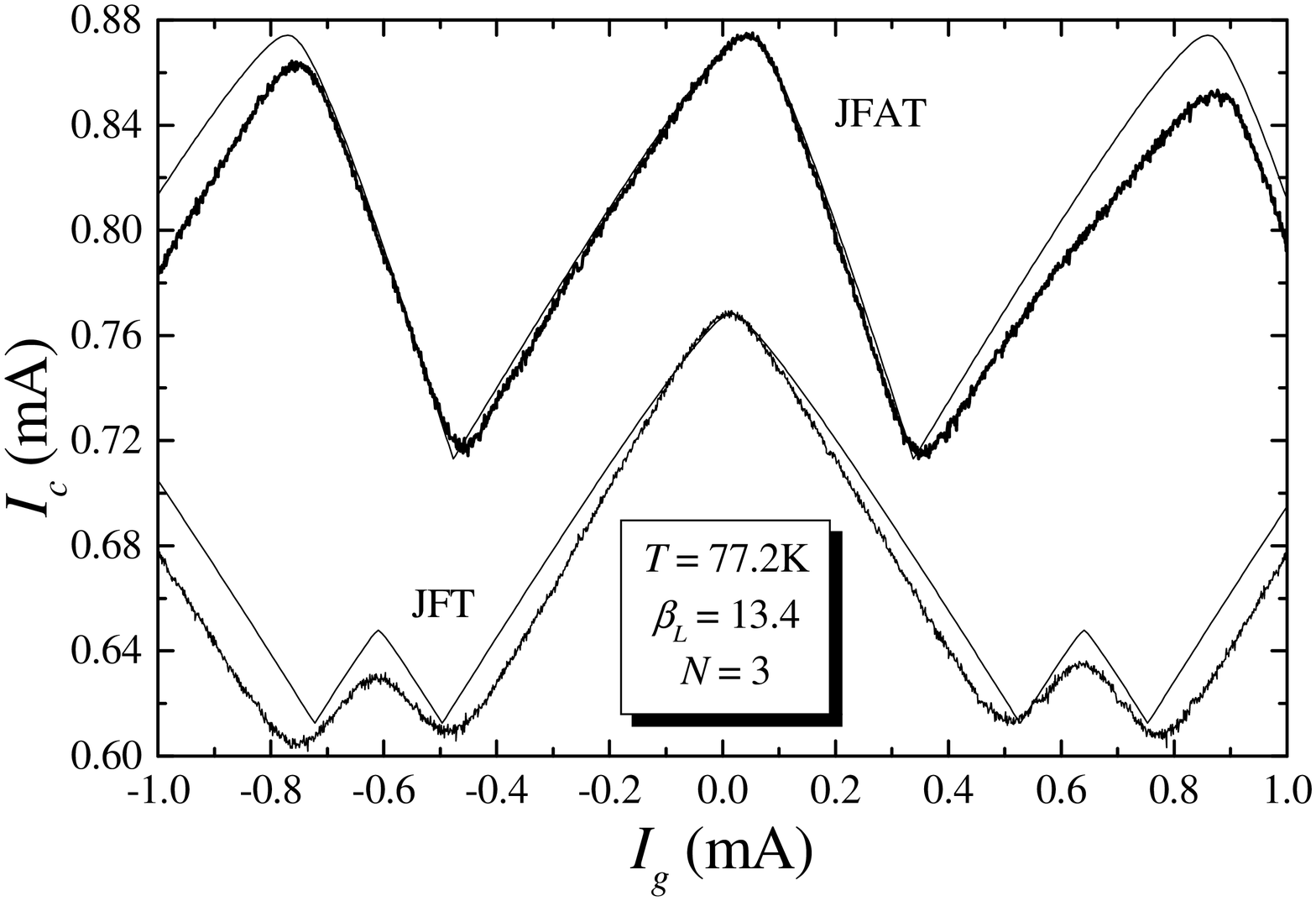}}           %
\caption[]{\small Measured $I_c(I_g)$-characteristics of a JVFT with $N$=3
operated as a JFAT and as a JFT. For comparison the calculated
characteristics are shown as thin lines. For clarity the JFT-data are offset
by $\Delta I_c\rm =-0.1mA$.                                   %
\label{f-cmp}}                                                        %
\end{figure}

Figure \ref{f-cmp} shows a direct comparison of the
simulated and the measured $I_c(I_g)$-characteristics
of the device shown in Fig.\ref{f-photo} in both modes
of operation at 77.2K. We find reasonably good
agreement of the simulated and the measured $I_c(I_g)$
characteristics, with $\beta_L = 13.4$ determined from
the modulation depth of $I_c$ \cite{gross95a,schuler98}.
Deviations are mainly caused by two aspects, which are
not considered in the algorithm. Firstly, the critical
current of the individual junctions is suppressed in a
magnetic field,  leading to reduced maximum $I_c$ at the
side maxima of the $I_c(I_g)$-curves as compared to the
maximum $I_c$ at $I_g=0$. Secondly, our algorithm does
not include thermal fluctuations, which lead to a
rounding of the measured $I_c(I_g)$-characteristics.

Evidently, the experimentally realized gain of the JFAT
($g_{meas}^{JFAT} = 0.643$) is substantially higher
than the gain realized in the JFT mode of operation
($g_{meas}^{JFT} = 0.386$), which can be attributed to
two effects: Firstly, a smaller period $\Delta I_g=4\pi
I_c^0(1+\frac{h_L}{w_L})/(k_i\beta_L)$ of the
$I_c(I_g)$-curves \cite{gross95a,schuler98} for the
JFAT increases its gain due to improved coupling to the
gate line. From the measured $\Delta I_g$ and with
$\beta_L=13.4$ we determine $k_i = 0.488$ for the JFAT
and $k_i = 0.326$ for the JFT. Secondly, an asymmetric
distortion of the $I_c(I_g)$ characteristics of the
JFAT increases its gain. This can be explained by
considering the JFAT as a parallel connection of two
asymmetric in-line type JFTs \cite{gross95a}, with
$(N-1)/2$ loops \cite{schuler98}. This interpretation
also explains the absence of the small intermediate
$I_c$-maxima for the 3-junction JFAT, which are present
for the 3-junction JFT [cf. Fig.\ref{f-cmp}].

\begin{figure}[bth]                                                   %
\center{\includegraphics [width=0.8\columnwidth] {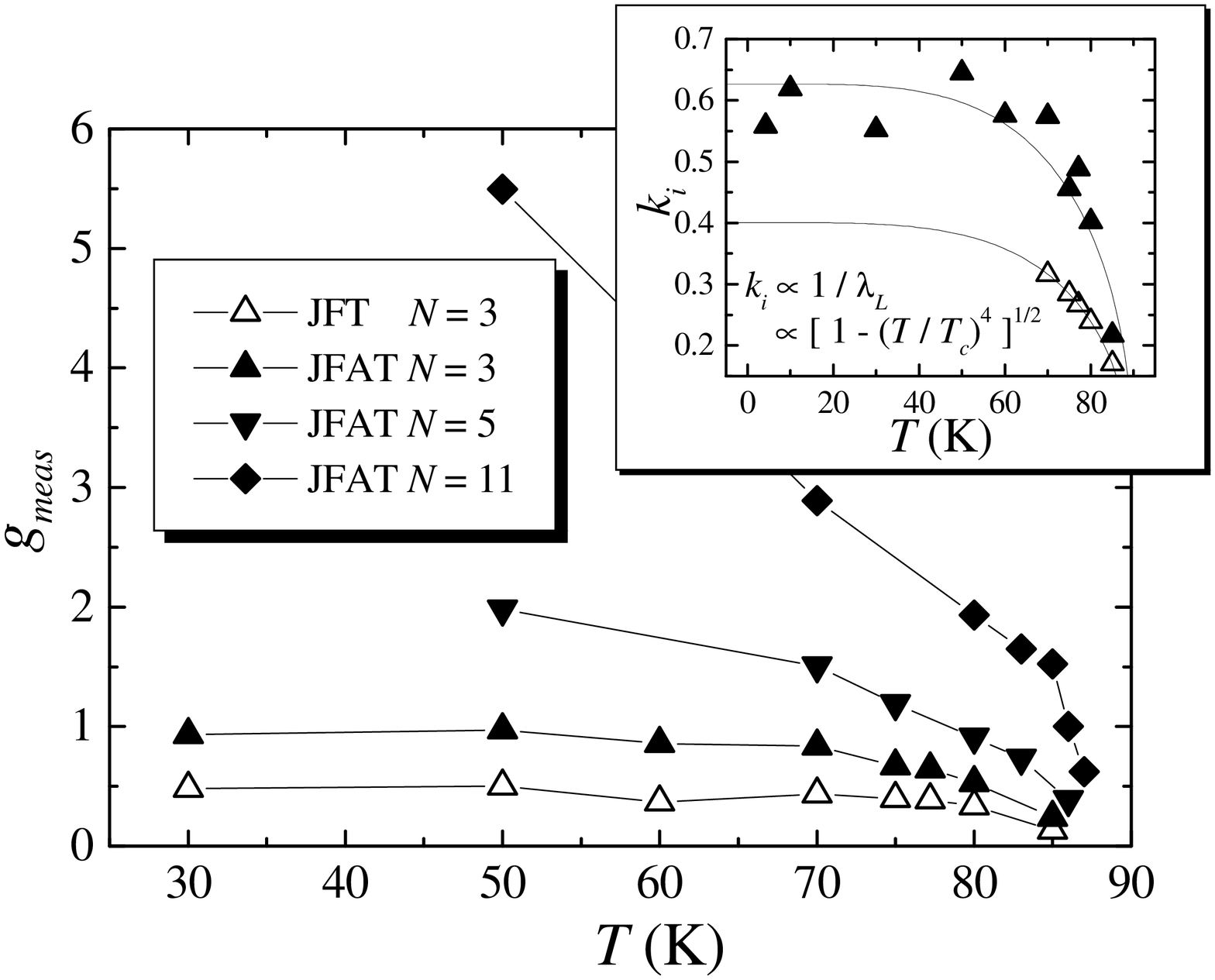}}            %
\caption[]{\small
Measured gain $g_{meas}$ vs. temperature $T$ for three devices. The inset
shows measured (symbols) and fitted (lines) coupling constants $k_i(T)$ for
the device with $N=3$ junctions operated as JFT and JFAT.
\label{f-gt}}                                                         %
\end{figure}

Figure \ref{f-gt} shows a compilation of measured gains and coupling
constants for different devices measured at variable temperature. We were
able to achieve gains as high as 5.5 for a device with $N=11$ junctions,
which was operated as a JFAT. Here, only gains have been regarded, where the
underlying $I_c(I_g)$-characteristic does not show a discontinuity, as in
such a point the gain cannot be defined at all.

\begin{figure}[b]                                                       %
\center{\includegraphics [width=0.8\columnwidth] {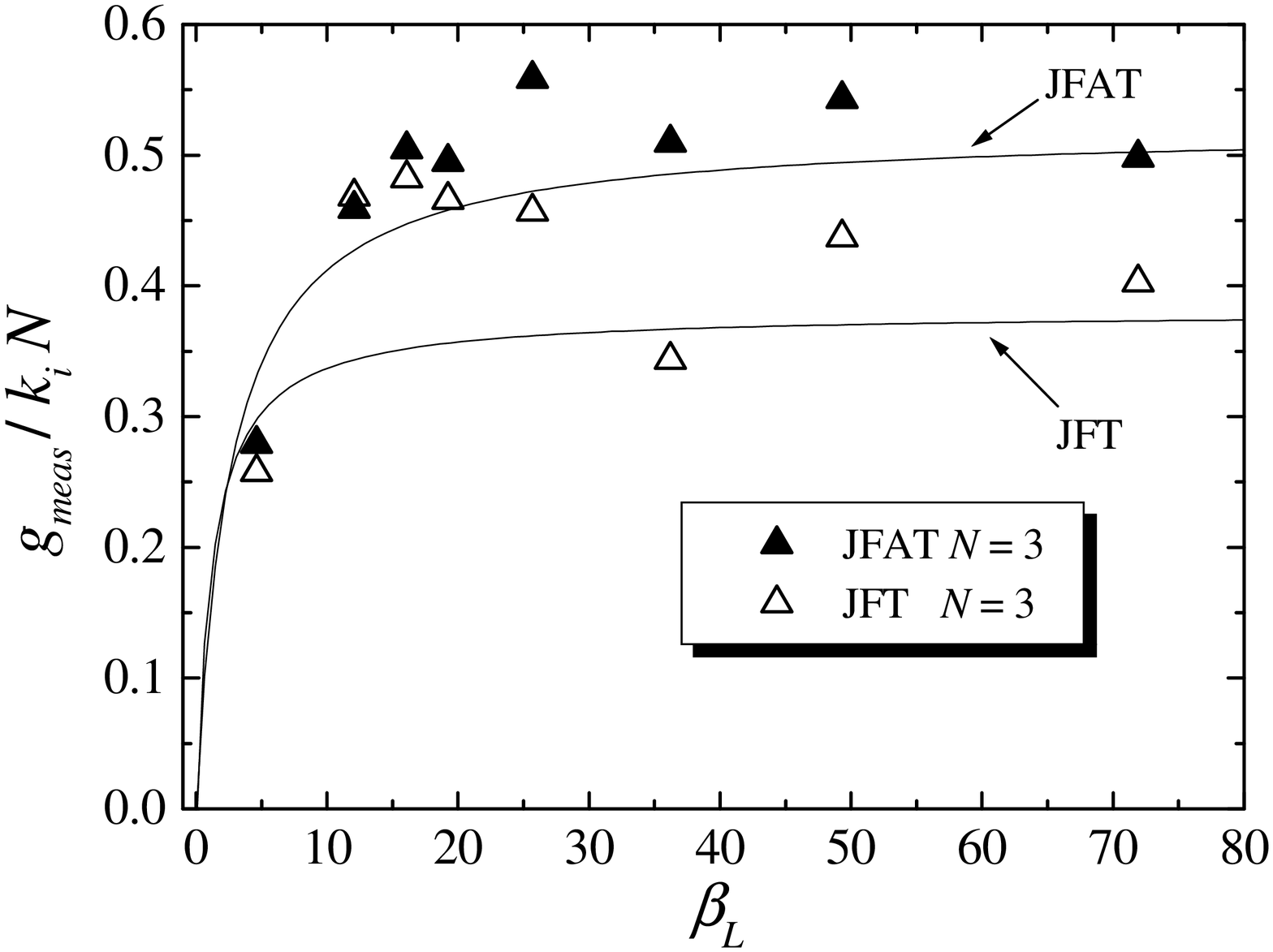}}            %
\caption[]{\small Measured normalized gain $g_{meas}/(k_i N$) vs. $\beta_L$
for a device with $N=3$ junctions operated as JFT and JFAT. The lines show
the theoretical prediction from eq.(\ref{e-fit}) based on numerical
simulations.                                                     %
\label{f-gnbl}}                                                         %
\end{figure}

We note that $g$ increases with decreasing $T$ for two
reasons: Firstly, we expect from our simulations that
$g$ increases with $\beta_L$, which becomes larger by
lowering $T$ due to increasing $I_c^0$. Secondly, the
gain increases linearly with $k_i$, which is expected
to approximately scale as $1/\lambda_L$
\cite{schuler98}, in fairly good agreement with the
data shown in the inset of Fig.\ref{f-gt}. To compare
the experimental data with the simulation results shown
in Fig.\ref{f-gsim} we normalize the measured gain by
the number $N$ of junctions and by the fitted $k_i(T)$
and plot it vs. $\beta_L$ as determined from the
modulation depth of $I_c$. The result is shown in
Fig.\ref{f-gnbl} for the device shown in
Fig.\ref{f-photo} with $N=3$.
We find reasonable agreement between the experimental data and the
theoretical lines, which are plotted from eq.(\ref{e-fit}) with the values
for the fitting parameters from table I. The significant scatter of the data
points is not surprising, given the scatter in the data of $k_i(T)$ for the
JFAT and the uncertainty in $k_i(T)$ for the JFT for $T<70$K [cf. inset of
Fig.\ref{f-gt}]. The most significant deviation from the numerical
simulation is found for the JFT-data which lie significantly above the
theoretical result for $10<\beta_L<25$. The reason for this deviation has
not been clarified yet.

In conclusion, our numerical simulations of the $I_c(I_g)$-characteristics
of JVFTs based on a parallel array of junctions with two different gate line
configurations predict superior behavior of the JFAT over the JFT with
respect to current gain. Our experiments on JVFTs from YBCO Josephson
junctions with cross gate geometry allow direct comparison of both types of
JVFTs which resulted in an experimental confirmation of the numerical
simulation results. The reason for the superior behavior of the JFAT is
twofold: (i) the JFAT can be viewed as two parallel JFTs with asymmetric inline
junction geometry which induces asymmetric $I_c(I_g)$-characteristics and,
hence, an increase in current gain over the symmetric JFT, and (ii) the
coupling between gate line and the junction array is almost a factor of two
higher in the JFAT configuration. The latter effect is not included in the
simulation results, but has been shown experimentally.


\begin{table}
\caption{Fit parameters obtained by fitting the simulation results to
eq.\ref{e-fit}.}
 \label{table1}
\begin{center}
\begin{tabular}{c|c|c|c}
\, \, Type \,\,  & \, \, $N$ \, \,       & \, \, \, $g_\infty$\, \, \,  & \,
\, \,  $\beta_0$ \, \, \,
\\ \hline JFT & 3 & 0.380 & 1.27
\\            &  $\geq 5$ & 0.390  & 0.80
\\ \hline
         JFAT &  3        & 0.521  & 2.64
\\            &  $\geq 5$ & 0.632  & 1.67
\end{tabular}
\end{center}
\end{table}

\end{multicols}               %

\begin{thebibliography}{99}

\bibitem{gross95a}
R. Gross, R. Gerdemann, L. Alff, T. Bauch,  A. Beck, O. M. Froehlich, D.
Koelle, A. Marx, Appl. Supercond.  {\bf 3}, 443 (1995).

\bibitem{gerdemann95}
R. Gerdemann, T. Bauch, O.M. Froehlich, L. Alff, A. Beck, D. Koelle, R.
Gross,
Appl. Phys. Lett. {\bf 67}, 1010 (1995).

\bibitem{berkowitz96}
S.J.~Berkowitz, Y.M.~Zhang, W.H.~Mallison, K.~Char, E.~Terzioglu, M.R.~Beasley,
Appl. Phys. Lett. \textbf{69}, 3257 (1996).

\bibitem{terzioglu96}
E.~Terzioglu, M.R.~Beasley, Y.M.~Zhang, S.J.~Berkowitz,
J. Appl. Phys. \textbf{80}, 5483 (1996).

\bibitem{schuler98}
J\"{u}rgen Schuler,
Diploma Thesis, Universit\"{a}t zu K\"{o}ln (1998).

\bibitem{mayer95}
B.~Mayer, H.~Schulze, G.M.~Fischer, R.~Gross,
Phys. Rev. B \textbf{52}, 7727 (1995).






\end{thebibliography}
\end{document}